\documentclass[reprint,prb,aps,twocolumn]{revtex4}

\usepackage[latin1]{inputenc}
\usepackage{amsmath}
\usepackage{amsfonts}
\usepackage{amssymb}
\usepackage{epsfig}
\usepackage{graphicx}
\usepackage{lscape}

\begin{document}
\title{
Brick by Brick Computation of the Gibbs Free Energy of Reaction in Solution 
Using Quantum Chemistry and COSMO-RS\\ 
}
\author{Arnim Hellweg and Frank Eckert}
\affiliation{COSMOlogic GmbH \& Co. KG,
Imbacher Weg 46,
D-51379 Leverkusen,
Germany}

\date{\today}

\begin{abstract}

The computational modelling of reactions is simple in theory but can be
quite tricky in practice. This article aims at the purpose of providing
an assistance to a proper way of describing reactions theoretically and
provides rough guidelines to the computational methods involved.

Reactions in liquid phase chemical equilibrium can be described 
theoretically in terms of the Gibbs free energy of reaction. This 
property can be divided into a sum of three disjunct terms, namely 
the gas phase reaction energy, the finite temperature contribution 
to the Gibbs free energy, and the Gibbs free energy of solvation. 
The three contributions to the Gibbs free energy of reaction can be 
computed separately, using different theoretico--chemical calculation 
methods. While some of these terms can be obtained reliably by 
computationally cheap methods, for others a high level of theory 
is required to obtain predictions of quantitative quality.

In order to propose workflows which can strike the balance between 
accuracy and computational cost, a number of benchmarks  
assessing the precision of different levels of theory is given. 

As an illustrative example, the low-temperature hydrogenation reaction 
of acetaldehyde to ethanol in solvent toluene is shown.

\end{abstract}
\maketitle

\section{Introduction}
\label{Introduction} 

In chemical reactions the change of the Gibbs free energy and thereby the 
equilibrium constants determines in which direction the reaction is driven. 
The knowledge of its amount and sign can be helpful for a better 
understanding of reactions in general, as well as for the analysis of 
reaction pathways and process optimization.\cite{basf,peters06}
With the power of modern computers and well-established computational procedures,
chemical equilibria nowadays can be computed routinely. Nevertheless, an open 
question that often remains is how well a chemical reaction can be predicted without 
the use of experimental data, and what effects and contributions need to be 
taken into account.

What makes this question complicated is the plain fact that actual real--life 
reactions take place at finite temperatures and quite regularly in solution. 
The correct description of these conditions calls for appropriate contributions 
to be included in the theoretical/computational description of a reaction. 
Reactions in solution typically are treated in terms of a thermodynamical cycle,
which computes the free energy contributions of all reacting species separately 
in the gas phase and in solution (see below). The connection of the gaseous and 
liquid phase in the thermodynamical cycle is described in terms of the compounds 
free energy of solvation $\Delta G_{solv}$. The other free energy contributions used
in the thermodynamic cycle are: the reaction energy in the gas phase 
$\Delta E_{gas}$, and the thermodynamics (finite temperature) contribution 
$\Delta G_{therm}$, which stems from rotational, translational, and vibrational 
degrees of freedom of the molecules involved.

The calculation of $\Delta E_{gas}$, the reaction energy in the gas phase, is rather 
straightforward. It is just the energy difference of the gas phase electronic energies of the 
product compounds and the reactant compounds. These can be computed by $ab~initio$ 
quantum mechanics for isolated molecules at zero Kelvin. In principle, the accuracy of 
this property is limited by the size of the involved molecules only, as the size 
determines the level of theory that can be applied to approximate the quantum mechanical
Schr{\"o}dinger equation of the molecule. The level of theory means the combination of
the quantum chemical method and the basis set. Both are 
approximations that can be used as parameters for improvement.\cite{TheBook,bak00} 

The finite temperature contribution to the Gibbs free energy 
$\Delta G_{therm}$ can also be determined from $ab~initio$ quantum mechanics.
Typically, the harmonic oscillator and ideal gas approximations 
are applied to obtain molar thermodynamic functions.\cite{GWP} 

While the prediction of gas phase reaction energies is a field where 
first principles quantum mechanics methods are applied,\cite{Friedrich15,GMTKN24} 
the computational prediction of the Gibbs free energy of solvation $\Delta G_{solv}$
usually is done with (semi-) empirical methods based on heuristic assumptions or regressions.
Among the prediction methods used are molecular mechanic force fields in 
molecular dynamic (MD) or Monte Carlo (MC) simulations,
quantitative structure--property relationships (QSPR),
COSMO-RS, group contribution methods, or neural networks
(for an overview and classification of such methods
see Ref. \onlinecite{solvnn} and references therein).

Thus, for the prediction of the overall Gibbs free energy of reaction in solution,
methods from both worlds - first principles quantum mechanics and more-or-less 
empirical prediction methods - have to be combined.
It has to be made sure that the expectable accuracies of the different 
prediction methods used in such a process are somewhat balanced, and that 
no errors are propagated in the course of the computations. Clearly, this task can 
be quite challenging for a user common to only one (or neither) of these worlds. 
This manuscript tries to be a rough guide to these worlds as it
tries to clarify some of the confusion that can arise about different $ab~initio$ 
methods or about the different ways to compute and add solvation effects.

\section{Computational procedures}
\label{procedures}

In this study we used the TURBOMOLE version 7.0\cite{TM} program package 
for the $ab~initio$ quantum mechanics calculations, and the COSMO$therm$ 
version C30-1701 implementation\cite{CT} of the COSMO-RS solvation model\cite{Eckert02,Klamt00,Klamt98} to 
obtain the free energy of solvation $\Delta G_{solv}$ of all reactant and product species 
involved in the reaction. 

The thermodynamic cycle describing a bimolecular reaction in solution can be sketched like this: 
\begin{widetext}
\begin{tabbing}
 \qquad \qquad \qquad \qquad   \= A$_{(gas)}$ + \=  B$_{(gas)}$ $\rightarrow$ \= ~ C$_{(gas)}$   \= ~:~ $\Delta G_{gas}$ ~~\=  = $\Delta E_{gas}+ \Delta G_{therm} $ \\
 \qquad \qquad\qquad \qquad   \> $\downarrow$ \> $\downarrow$ \> ~ $\downarrow$  \> ~:~  $ \Delta G_{solv}$  \> \\
 \qquad \qquad \qquad \qquad   \> A$_{(soln)}$ + \> B$_{(soln)}$ $\rightarrow$ \> ~ C$_{(soln)}$  \> ~:~ $\Delta G_{soln}$ \> = $\Delta E_{gas} + \Delta G_{therm} +\Delta G_{solv} $ \\
\end{tabbing}
\end{widetext}

The free energy contributions $\Delta E_{gas}$, $\Delta G_{therm}$, and $\Delta G_{solv}$ 
of the reactant species A, B, and product C, depend 
on the properties of the reactants and the products. Each one
has to be computed independently on the same level of theory. This way 
the resulting $\Delta G$ and $\Delta E$ energy differences are additive.

\begin{equation}
\Delta E_{gas}   =  E_{gas}(C)-( E_{gas}(A)+ E_{gas}(B) ) 
\end{equation}
\begin{equation}
\Delta G_{therm} =  G_{therm}(C) - ( G_{therm}(A)+ G_{therm}(B) ) 
\end{equation}

Thus, the full cycle can be divided into three autonomous steps, 
for which different levels of theory can be applied.

\subsection*{Quantum chemical gas phase energy $\Delta E_{gas}$}

In order to obtain electronic energies from quantum mechanics, 
it is crucial to have a valid structure (3D-geometry) of the molecules.  
In the best and most well--defined case, the energy is the 
one of the minimum energy structure from an geometry optimization obtained 
on the same level of theory. 
However, optimizations on accurate levels may become computationally 
expensive, while the structural improvements compared to 
optimizations on more economic quantum mechanics levels are often small.
Due to this observation, it is common practice to combine 
optimizations on low--cost levels with a single--point 
calculation on an accurate level.


\subsection*{Thermodynamic Gibbs free energy $\Delta G_{therm}$}

For the computation of $\Delta G_{therm}$ of a molecule, its
structure has to be optimized and the vibrational frequencies
have to be evaluated. These calculations have to be performed 
on the same level of theory, as vibrational frequencies
are only defined at stationary points (i.e. global or local minima, 
or saddle points on the molecules potential energy surface).
The evaluation of vibrational frequencies is computationally
quite demanding, but fortunately, low--cost methods already yield
results of sufficient quality (see below).

\subsection*{Free energy of solvation $\Delta G_{solv}$}

Solvation free energies can be computed with the COSMO-RS 
solvation model. They can be added individually to each component of the cycle. 
COSMO-RS is a prediction method for thermodynamic properties of liquids that is based
on surface charge descriptors as provided by quantum mechanics calculations, 
but also involves a small number of fitted parameters specific to the 
quantum mechanics level used.\cite{Eckert02,Klamt00,Klamt98} 
See below for more details on COSMO-RS. 

Generally, it is not necessary to use the same computational level for 
the COSMO-RS calculation of $\Delta G_{solv}$ and the $ab~initio$ 
quantum mechanics level used for $E_{gas}$ and $G_{therm}$. As COSMO-RS
is adjusted to specific quantum mechanics levels it is sufficient to
use computationally cheap density functional (DFT) methods. 
The application of COSMO-RS in chemical and engineering thermodynamics 
(e.g. prediction of binary VLE or LLE data, activity coefficients in solution, or vapor pressures) 
and particularly in reaction modeling typically requires a high quality of  
property predictions of mixtures of small to medium sized molecules 
(up to 25 non--hydrogen atoms). The specific methods used in COSMO-RS were 
chosen according to the prediction quality of the surface charge descriptors 
used in COSMO-RS. Currently three levels are commonly used and suitable for the 
purpose of $\Delta G_{solv}$ prediction in reaction modeling (see below for 
definitions of the quantum mechanics levels and basis sets used):

\texttt{ BP-TZVP }\cite{Eckert02}: This level is considered to one of the two "high quality" 
working levels currently offered in COSMO-RS. It uses a COSMO charge surface calculation on a 
geometry optimized on BP DFT with def-TZVP basis set. 
The \texttt{BP-TZVP} level is available in several 
quantum mechanics program suites, such as 
TURBOMOLE\cite{TM}, Gaussian\cite{Gaussian}, GAMESS\cite{Klamt97}, 
ORCA\cite{ORCA}, MOLPRO\cite{MOLPRO}, Q-Chem\cite{QCHEM}, and some more.

\texttt{ DMOL3-PBE }\cite{Klamt00,Klamt98}: This level is considered to one of the two "high quality" 
working levels currently offered in COSMO-RS. It uses surface charge of a 
PBE DFT optimized structures with numerical DNP basis set. 
The \texttt{DMOL3-PBE} level is available in the DMOL3\cite{DMOL3}
quantum mechanics program.

\texttt{ BP-TZVPD-FINE }\cite{FINE}: This level is considered to be the "best quality"
calculation method that is currently offered in COSMO-RS. It uses a single--point
COSMO surface charge calculation on BP DFT with def2-TZVPD basis set
and a smooth radii--based isosurface cavity,
which is done upon a geometry optimized on BP DFT with def-TZVP basis set and COSMO.
The \texttt{BP-TZVPD-FINE} level currently is available in the 
TURBOMOLE\cite{TM} quantum mechanics program.

As the \texttt{BP-TZVPD-FINE} level currently is considered to be the best quality 
method available within 
COSMO-RS. It is highly recommended to use this level for all predictions of $\Delta G_{solv}$.

\subsection*{Free energy of reaction in solution $\Delta G_{soln}$}

The Gibbs free energy for a reaction in solution can be written
\begin{equation} \label{e0}
\Delta G_{soln} = \Delta E_{gas} + \Delta G_{therm} + \Delta G_{solv}.
\end{equation}

Each of the three contributions or bricks needed can be decomposed into the time--critical quantum chemical steps, denoted as QM levels the Eqs. (\ref{e1},~\ref{e2},~\ref{e3}). Optional levels are given in parentheses. 

\begin{widetext}

\begin{equation} \label{e1}
\begin{aligned}
\Delta E_{gas}  \subseteq \{
\underbrace{\text{geometry optimization}}_{\substack{\text{QM level 1}
    \\ \text{gas} }} \underbrace{(, \text{single point energy})}_{\substack{\text{QM level 2} \\ \text{gas} }} \}\\ 
\end{aligned}
\end{equation}

\begin{equation} \label{e2}
\begin{aligned}
\Delta G_{therm}  \subseteq \{
\underbrace{\text{geometry optimization ,  frequencies}}_{\substack{\text{QM level 3} \\ \text{gas} }}  
  , \text{thermochemistry}  \}\\ 
\end{aligned}
\end{equation}

\begin{equation} \label{e3}
\begin{aligned}
\Delta G_{solv}  \subseteq \{
\underbrace{\text{geometry optimization}}_{\substack{\text{QM level  4}\\ \text{gas and COSMO}}} 
     \underbrace{(,\text{single point energy})}_{\substack{\text{QM level 5}\\ \text{gas and COSMO}}}  
     , \text{COSMO-RS} \}
\end{aligned}
\end{equation}

\end{widetext}

\section{Methods}
\label{methods}

\subsection*{Quantum Chemical Methods}

Quantum mechanics is a real zoo of methods. We restrict us here to 
single--reference methods from the fields of $ab~initio$ wave function 
theory (WFT) and of density functional theory (DFT).

As WFT methods we used 
Hartree--Fock (HF), second--order M{\o}ller--Plesset perturbation theory (MP2), 
spin--component scaled MP2 (SCS-MP2), coupled--cluster singles--and--doubles 
(CCSD), and CCSD with a perturbative correction for connected triples (CCSD(T)).
With these methods systematic improvements are possible,\cite{TheBook}
but they are only applicable, if the HF reference wave function is a good
approximation.
CCSD is considerably more expensive than MP2 but quite regularly not superior for certain properties.
In particular, geometries and reaction energies are usually better on MP2 level.
The SCS-MP2 approach is not fully $ab~initio$ as it includes two empirical parameters. 
However, it often performs better than regular MP2 at the same computational
costs. It can be recommended to use it in standard applications.\cite{SCS-MP2}

For the DFT studies we used density functionals of different classes.
As prototype for functionals in the generalized gradient 
approximation (GGA) BP,\cite{B88,P86} and  PBE\cite{PBE} were applied,
for hybrid-GGA (HGGA) B3LYP,\cite{b3lyp93} and PBE0\cite{PBE0},
for meta-GGA (MGGA)   TPSS,\cite{tpss} and M06-L\cite{M06-L},
for hybrid-MGGA (HMGGA)  M06,\cite{M06} and PW6B95\cite{PW6B95},
and for GGA plus dispersion correction with Becke-Johnson (BJ) damping\cite{BJ-damp}
(GGA-D3(BJ)) B97-D.\cite{B97-D}

The pure GGA functionals like BP, PBE, or B97-D are nearly linear scaling and 
by far the fastest methods if they are combined with the MARI-J approximation\cite{marij}. 
The BP and PBE functionals hold almost no empirical assumptions, are very robust, 
and usually perform well over the whole of the periodic system of elements. 
Augmenting DFT with the empirical dispersion corrections by Grimme\cite{BJ-damp}
comes at no additional computational cost. In connection with the functional B97-D 
good geometries can be obtained, in particular for larger molecules.\cite{B97-D}
The B3LYP hybrid functional is extremely popular among chemists. 
It is well--known for yielding very good results for geometries of organic molecules.\cite{dft-papers}
The PBE0 functional is often a good choice for transition barrier height 
as well as for excited state calculations.\cite{adamo09}
TPSS and M06-L are computationally only slightly more demanding than pure GGAs. 
They are known to give good results for metals, transition metals, and inorganic systems. \cite{3dTM} 
M06 includes non--covalent interactions and was developed for main group 
and transition metal thermochemistry or organometallics.\cite{M06}
The PW6B95 fuctional often is a top performer for geometries, reaction energies,
and thermochemistry in benchmarks of organic and main group chemistry.\cite{GMTKN30}
There are literally tons of papers available that try to assess the performance 
of density functionals, and many people have very strict opinions about it.
One should keep in mind that it could very well be, that there is no functional that works in every case.  
So, applying an older, well--established functional in a new investigation
could yield in less surprises than a very recent, but not fully tested one.

\subsection*{Basis Sets}

Basis sets for quantum mechanics calculations are another zoo.
We restrict ourselves here to the Karlsruhe segmented contracted 
Gaussian basis sets of Ahlrichs an co-workers.
Popular other choices could be the Pople-style basis sets\cite{Ditchfield71,Hehre:72,Binkley80a,Krishnan:80}  
or the series of Dunning and co-workers.\cite{Dunning:89,Kendall:92,Woon:93,Woon:94b}

The applied basis set are of 
split--valence (SV), triple--$\zeta$ (TZV), and quadruple--$\zeta$ 
(QZV) valence qualities.\cite{Sch92,Sch94,Wei03,Wei05}
These classes are optimized for a certain quality throughout the periodic table.
A suffix P means polarized and PP heavily polarized. 
The def- or def2- prefixes are used to distinguish which effective core 
potentials (ECPs) are applied to elements from Rb to Rn. 
The more recent def2-sets also have some improve polarization functions. 
For QZVPP the def and def2 are equal.
Additional diffuse basis functions are needed for computations involving 
anions (e.g. the prediction p$K_a$ values), or molecular properties related 
to electron densities that are not localized in the valence region.
Diffuse basis functions are denoted by suffix D.

The resolution--of--the--identity (RI) approximation
is used for the SCF\cite{Whitten:JCP58,Dunlap:JCP71,Vahtras:93, marij, Feyereisen:93} 
as well as the MP2 and coupled--cluster calculations.\cite{Feyereisen:93,Wei98} 
The exploitation of the RI approximation reduces the computational 
costs by an order of magnitude without the loss of accuracy, but
it requires additional so-called auxiliary or fitting basis sets.
\cite{Eic97,Wei06:jbasen, Wei98,Hae05, Hell07}

In general, DFT is less susceptible to the choice of basis sets than
WFT. Nevertheless choosing appropriate basis sets from the beginning 
reduces the risk of bothering with inconsistencies and inexplicable artifacts later on.
In DFT studies basis set of already SV quality often yield good results for 
structures, for an energetic assessment TZV quality should be used.
In WFT studies basis below TZV quality should only be used for preliminary, 
explorational investigations.
When in doubt, it may be useful to check if basis set convergence is reached: 
simply try to use the next larger basis set in the calculation and check if 
the deviations of the target property vanishes for the different basis set levels.
Sometimes a reduced--size model system might be needed, if the calculation 
gets too demanding for the actual molecule in question.

In the present study, we applied DFT fine grids (m4 in TURBOMOLE notation) 
and tight convergence 
criteria, except for the COSMO calculations where we used the default settings.

\subsection*{Solvation Models}

In this work we are only considering the COSMO-RS method as prediction tool for the 
free energy of solvation $\Delta G_{solv}$. COSMO-RS is a theory of interacting molecular surfaces 
as computed by quantum mechanics methods. COSMO-RS combines an electrostatic theory of locally 
interacting molecular surface descriptors (which are available from QM calculations) with a 
statistical thermodynamics methodology. 

\begin{figure}
\begin{center}
\includegraphics[width=3.0cm,angle=0]{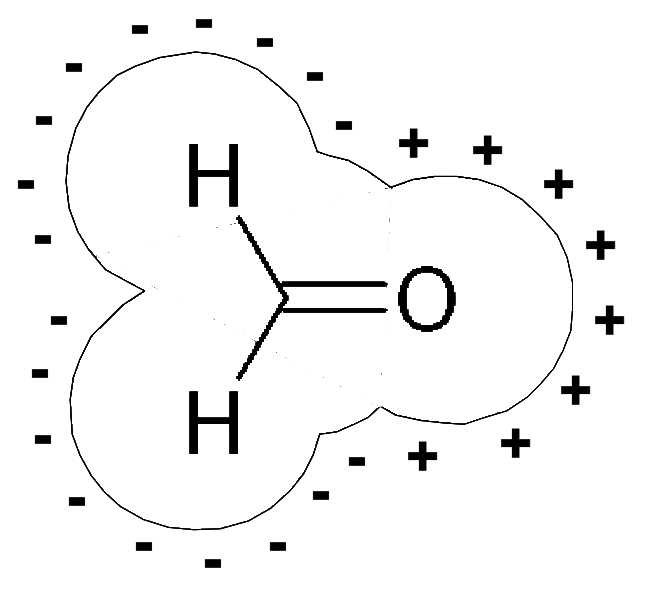} 
\end{center}
\caption{
Thumb-sketch on a cavity for quantum chemical continuum solvation models.
\label{fig.cavity}}
\end{figure}

The quantum chemical basis of COSMO-RS is COSMO, the COnductor-like Screening MOdel,\cite{COSMO} which belongs 
to the class of quantum chemical continuum solvation models (CSMs). In general, basic quantum chemical 
methodology describes isolated molecules at a temperature of T=0 K, allowing a realistic description 
only for molecules in vacuum or in the gas phase. CSMs are an extension of the basic QM methods towards 
the description of liquid phases. CSMs describe a molecule in solution through a quantum chemical 
calculation of the solute molecule with an approximate representation of the surrounding solvent as a continuum. 
Either by solution of the dielectric boundary condition or by solution of the Poisson-Boltzmann equation, 
the solute is treated as if embedded in a dielectric medium via a molecular surface or cavity that is 
constructed around the molecule. In Fig.~\ref{fig.cavity} an example of a cavity is sketched.
Hereby, normally the macroscopic dielectric constant of the solvent is used. 
COSMO is a quite popular model based on a slight approximation, which in comparison to other CSMs achieves 
superior efficiency and robustness of the computational methodology. The COSMO model is available in many 
quantum chemistry program packages such as TURBOMOLE\cite{TM}, Gaussian\cite{Gaussian}, GAMESS\cite{Klamt97}, 
ORCA\cite{ORCA}, MOLPRO\cite{MOLPRO}, and Q-Chem\cite{QCHEM}. However, as has been shown elsewhere,\cite{Klamt06} 
that the continuum description of CSMs is based on an erroneous physical concept. In addition, concepts of 
temperature and mixture are missing in CSMs.

\begin{figure*}[!ht]
\begin{center}
\includegraphics[width=8.9cm,angle=0]{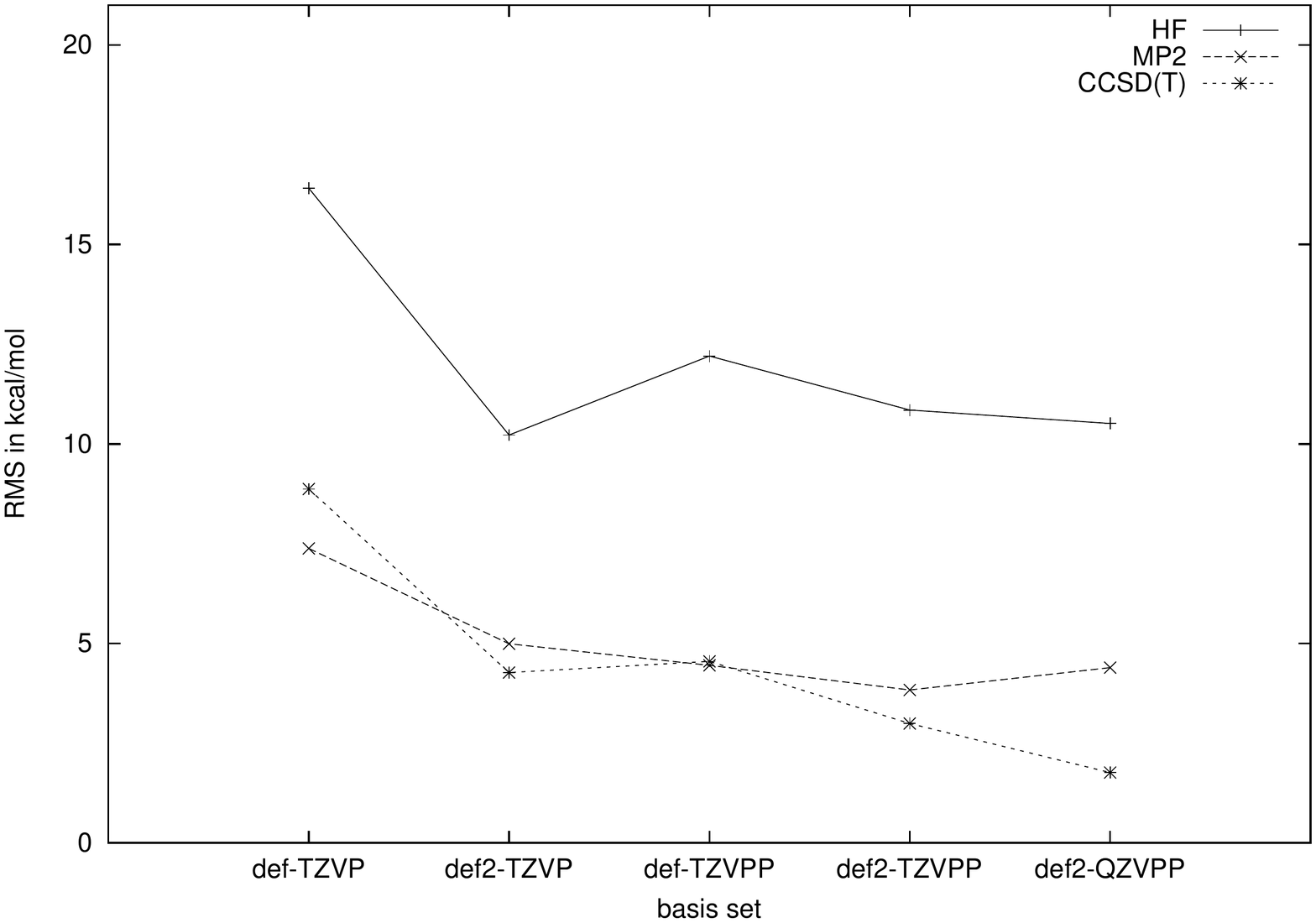} 
\includegraphics[width=8.9cm,angle=0]{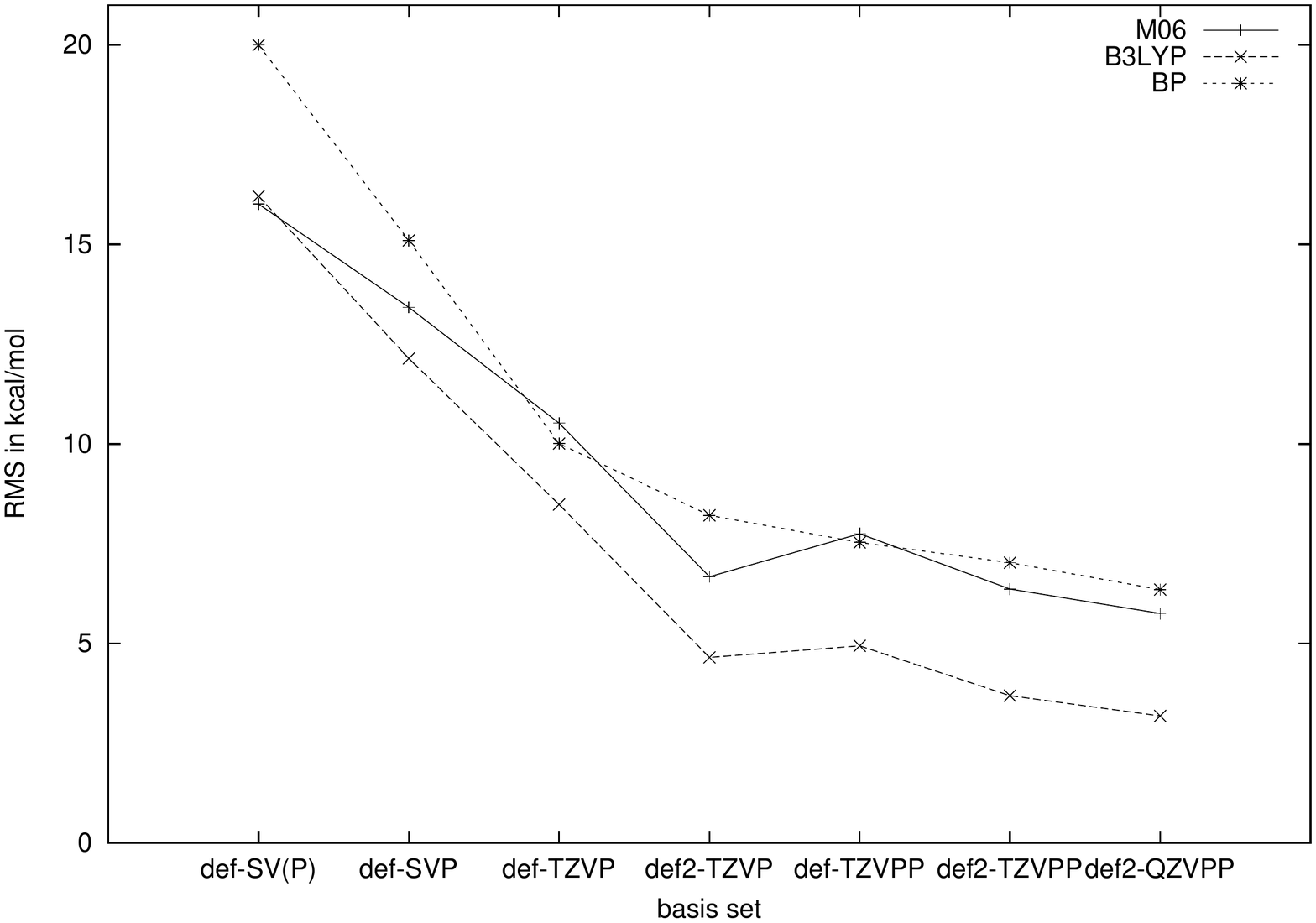}
\end{center}
\caption{
The basis set convergence of reaction energies $\Delta E_{gas}$ of WFT (left) and DFT (right) methods.
The RMS error between calculated and experimental values from the GMTKN24/G2RC set is given in kcal/mol.
\label{fig.error-gas-wft}}
\end{figure*}

COSMO-RS, the COSMO theory for "real solvents"\cite{Eckert02,Klamt00,Klamt98,Klamt97} goes far beyond simple CSMs 
in that it integrates concepts from quantum chemistry, dielectric continuum models, electrostatic surface interactions 
and statistical thermodynamics. Still, COSMO-RS is based upon the information that is evaluated by QM-COSMO calculations. 
Basically QM-COSMO calculations provide a discrete surface around a molecule embedded in a virtual conductor.\cite{COSMO} 
Of this surface each segment $i$ is characterized by its area $a_{i}$ and the screening charge density (SCD) $\sigma_i$ 
on this segment. The SCD takes into account the electrostatic screening of the solute molecule by its surrounding 
(which in a virtual conductor is perfect screening) and the back-polarization of the solute molecule. 
Within COSMO-RS theory a liquid now is considered an ensemble of closely packed ideally screened molecules. 
Each piece of the molecular surface is in close contact with another one. Assuming that there still is a conducting 
surface between the molecules, i.e. that each molecule still is enclosed by a virtual conductor, in a contact area the 
surface segments of both molecules have net SCDs. In reality there is no conductor between the surface contact areas. 
Thus an electrostatic interaction arises from the contact of two different SCDs. Hydrogen bonding (HB) can also be 
described by the two adjacent SCDs. In addition, dispersion interactions are taken into account. 
The link between the microscopic surface interaction energies and the macroscopic thermodynamic properties of a 
liquid is provided by statistical thermodynamics. Since in the COSMO-RS view all molecular interactions consist of 
local pair wise interactions of surface segments, the statistical averaging can be done in the ensemble of interacting surface pieces. 
Such an ensemble averaging is computationally efficient - especially in comparison to the computationally very demanding 
molecular dynamics (MD) or Monte Carlo (MC) approaches which require averaging over an ensemble of all possible different 
arrangements of all molecules in a liquid. To describe the composition of the surface segment ensemble with respect to the interactions 
(which depend on SCDs only), only the probability distribution of the surface charges has to be known for all compounds $i$.
Please consider the original COSMO-RS theory papers\cite{Eckert02,Klamt00,Klamt98,Klamt97} for more information on this.

The majority of larger and more complex compounds can be existent in more than one conformation, which means that they have relevant 
metastable energy minima in addition to the global energy minimum. Fortunately, the conformational ambiguity can be disregarded 
in many cases for the calculation of chemical potentials and phase equilibria with COSMO-RS. This is the case if the SCDs of the different 
conformations are very similar, as e.g. for bond--rotation conformations in alkane chains. In such cases the thermodynamic equilibria are 
unaffected by the conformational ambiguity, and the compound can be well described by its minimum energy conformation. If however, 
the polarity of the conformations is very different, in particular if intramolecular hydrogen bond is possible in the molecule, 
the free energy difference may change strongly between a polar solvent such as water and a non-polar solvent or the gas phase. 
In this case different molecular conformations have to be taken into account in COSMO-RS. A compound $i$ can be represented by a 
set of SCDs for the conformers. The population of a conformer $j$ in solvent $S$ is calculated according to the Boltzmann 
distribution between states of different free energy ($G_{j}^{S} = E_{j}^{COSMO} + \mu_{j}^{S}$). 

\begin{table*}[!ht]
\caption{The RMS error of $\Delta E_{gas}$ calculated by electron correlation methods with different basis sets compared to experimental values from the GMTKN24/G2RC set in kcal/mol. 
With extra. CCSD(T) the values extrapolated by means of Eq. \ref{extra} are denoted.
\label{tab.e-error-wft} }
\smallskip
\begin{tabular}{lcccc}  \hline
basis set               & CCSD(T)         & CCSD  &  SCS-MP2 &  MP2    \\ \hline
def2-QZVPP              &  1.76           &  3.83 &  2.51    &  4.39   \\
def2-TZVPP              &  2.99           &  4.24 &  2.74    &  3.83   \\
def-TZVPP               &  4.55           &  5.92 &  3.65    &  4.45   \\
def2-TZVP               &  4.27           &  3.42 &  4.84    &  4.99   \\ \hline
                        &  extra. CCSD(T) &       &          &         \\
def2-QZVPP/def2-TZVPP   &  1.75           &       &          &         \\
def2-QZVPP/def-TZVPP    &  1.72           &       &          &         \\
def2-QZVPP/def2-TZVP    &  1.74           &       &          &         \\ \hline
\hline \noalign{\smallskip}
\end{tabular}
\end{table*}

Thus, COSMO-RS is able to compute macroscopic thermodynamic properties of liquids, such as the free energy of solvation $\Delta G_{solv}$, 
with the help of quantum mechanics derived descriptors. COSMO-RS depends on an extremely small number of adjustable parameters 
(seven basic parameters plus nine dispersion parameters) some of which are physically predetermined. COSMO-RS parameters are 
not specific of functional groups or molecule types. The parameters have to be adjusted for the QM-COSMO method that is used 
as a basis for the COSMO-RS calculations only. Hence, the resulting parametrization is completely general and can be used to 
predict the properties of almost any imaginable compound mixture or system. In this work the COSMO$therm$\cite{CT} implementation 
of COSMO-RS was used to compute all thermodynamic properties relevant to the transition from the quantum mechanical gas phase
computation to the liquid phase.

The free energy of solvation $\Delta G_{solv}$ for the reference state of 1 bar gas and 1 mol liquid is computed as the difference
of COSMO-RS predicted chemical potentials of the given compound $i$ in the gas phase and at infinite dilution in a liquid solvent $S$: 
\begin{equation} \label{gsolv-bar-mol}
\begin{aligned}
\begin{split}
 \Delta G_{solv,i} &= \mu_{i}^{S,\infty} - \mu_{i}^{gas}
\end{split}
\end{aligned}
\end{equation}
This $\Delta G_{solv}$ in the bar to mol reference frame is the one used in reaction $\Delta G_{soln}$ calculations.
The free energy of solvation in the more commonly used molar reference state of 1 mol/l gas phase and 1 mol/l liquid phase 
can bei obtained from the above with the help of the solvent density $\rho_{S}$, the solvent molar weight $MW_{S}$, and 
the molar volume of the ideal gas $V_{IG}$.
\begin{equation} \label{gsolv-molar}
\begin{aligned}
\begin{split}
 \Delta {G}_{solv,i} &= \mu_{i}^{S,\infty} - \mu_{i}^{gas} - RT ln( \frac{\rho_{S} V_{IG}}{MW_{S}} )
\end{split}
\end{aligned}
\end{equation}
A property closely related to $\Delta G_{solv}$ is the Henry law coefficient $k_{H}$. 
$k_{H}$ in turn, can be expressed as product of a compounds infinite dilution
activity coefficient $\gamma_{i}^{S,\infty}$ and its pure compound vapor pressure $p_{i}^{0}$.
\begin{equation} \label{khenry}
\begin{aligned}
\begin{split}
 \Delta k_{H} &= \{ \mu_{i}^{S,\infty} - \mu_{i}^{gas} \} \/ RT = \gamma_{i}^{S,\infty} p_{i}^{0}
\end{split}
\end{aligned}
\end{equation}
This opens a pathway for a further improvement of the predictions of $\Delta G_{solv}$. 
Usually COSMO-RS is used to predict both chemicals potentials in liquid and in gas $\mu_{i}^{S,\infty}$ and $\mu_{i}^{gas}$,
corresponding to the prediction of both infinite dilution activity coefficient $\gamma_{i}^{S,\infty}$ and 
pure compound vapor pressure $p_{i}^{0}$. If the pure compound vapor pressure $p_{i}^{0}$ is available experimentally,
it is possible to use this experimental value to scale and improve the $\Delta G_{solv}$ prediction via 
Eqs. \ref{khenry} and \ref{gsolv-bar-mol}. Typically, a small to medium improvement of about 0.0 - 0.3 kcal/mol 
can be achieved by the usage of experimental pure compound vapor pressures in free energy of solvation predictions.

\section{Benchmark}
\label{benchmark}

We benchmark the three bricks that build up the Gibbs free energy of 
reaction in solution separately to ensure that the performance of the 
sum of these is neither biased by error compensation nor error amplification.

\subsection{$ \Delta E_{gas}$}

Goerigk and Grimme set up the GMTKN24\cite{GMTKN24} database, which involves 
the G2RC subset of 25 experimental values for gas phase reaction energies of 
selected G2/97\cite{G2-97} small closed--shell molecules.  
We use this set to assess the accuracy of $\Delta E_{gas}$ calculations.



In Fig.~\ref{fig.error-gas-wft} the root mean squared (RMS) error of calculated
and experimental values in kcal/mol given for different methods and basis sets.
The errors of HF are by far the largest and this level of theory can not
be recommended for computations, yet it can be interesting to inspect its 
basis set convergence. 
A performance difference between the def- and def2-basis set for
WFT methods is notable here. 
The def2-basis performs better due to the improved polarization 
functions for the elements Al to Ar. 
On DFT level the effect is less pronounced in this test set, but
if heavier elements in connection with ECPs are used, the
def2 sets are also expected to give a better description than the def sets.

CCSD(T) calculations with small basis sets are a waste of time,
because in this case the accuracy is not better than for much cheaper methods.
CCSD(T) calculation with large basis sets are, however, very expensive or not feasible at all. 
One way out of this dilemma is to apply a small--to--large basis extrapolation scheme
based on the difference of MP2 and CCSD(T) energies.
This approach was applied to the determination of non-covalent
interactions by Hobza and co-workers,\cite{estCCSD(T)2002,estCCSD(T)2011} 
but recently it has also been benchmarked for reaction energies.\cite{Friedrich15} 
\begin{equation} \label{extra}
\begin{aligned}
\begin{split}
 E[CCSD(T)/big] &= E[MP2/big] \\ &+ E[CCSD(T)/small] - E[MP2/small]
\end{split}
\end{aligned}
\end{equation}
In Table~\ref{tab.e-error-wft} the accuracy of correlated methods and this extrapolated CCSD(T) 
approach is shown. It can be seen that the extrapolation scheme performs very well.
From the pure methods the SCS-MP2 works best, while it has the same computational cost
as MP2.

\begin{table}
\caption{The RMS error in kcal/mol of $\Delta E_{gas}$ for different functionals with def2-TZVP
compared to experimental values from the GMTKN24/G2RC set.
\label{tab.e-error-dft} }
\smallskip
\begin{tabular}{lcc}  \hline
        &  RMS  & functional class\\ \hline
BP      &  8.21 & GGA    \\
PBE     &  9.37 & GGA    \\ \hline
B3LYP   &  4.65 & HGGA   \\
PBE0    &  8.70 & HGGA   \\ \hline
TPSS    & 10.53 & MGGA   \\
M06-L   &  9.04 & MGGA   \\ \hline
M06     &  6.67 & HMGGA  \\
PW6B95  &  5.11 & HMGGA  \\ \hline
B97-D   &  7.90 & GGA-D3(BJ) \\ \hline
\hline \noalign{\smallskip}
\end{tabular}
\end{table}

The performance of different density functionals is quite similar if  
reasonable basis sets are used, see Table~\ref{tab.e-error-dft}. 
B3LYP is a positive outlier on this benchmark set. This may very well be due to the
fact that the parameters of the B3LYP functional were fitted on the G2 set of molecules 
- and the G2RC set used here is a subset of this dataset.
However, if larger molecules are investigated, the energetic characterization of B3LYP
was not found to be particularly improved in comparison with other functionals.\cite{G300,grim05}


\subsection{$ \Delta G_{therm}$}

In order to benchmark the accuracy of $\Delta G_{therm}$ computations,
we used the G2/97 set,\cite{G2-97} which consists of
55 small, mainly organic molecules.
As we are not aware of the existence of reliable experimental data 
for this property, we used CCSD(T)/def2-QZVPP computations as reference 
values. This level of theory should be close to methodical convergence 
and basis set limit. 

It has been argued that the application of empirical scaling factors to
vibrational frequencies can improve the prediction quality of 
IR spectra and ZPE.\cite{Alecu10}  
However, we did not apply scaling factors in the current work, 
because such scaling factors are different for each level of theory and 
not every combination of method and basis set is covered. Moreover, as 
the fit of the scaling factors depends on experimental data, and different
fit sets were used for different method and basis set combinations, the final
prediction quality might vary and hence does not allow for an unbiased 
comparison of the underlying DFT functionals or WFT methods and basis sets.

For an evaluation of $\Delta G_{therm}$, one has to keep in mind that the 
ZPE and the vibrational partition function are needed. It is crucial that 
these properties are calculated on a stationary point of the molecule's 
potential energy surface for the given level of theory. 
This means one has to optimize the geometry and run a subsequent frequency calculation
with the same functional or method and basis set.

\begin{table}
\caption{The basis set convergence on the basis of the RMS error in 
kcal/mol of $\Delta G_{therm}$ for WFT and DFT methods. 
\label{tab.therm-error-bs} }
\smallskip
\begin{tabular}{lccccc|cccc}  \hline
basis set    & CCSD(T) & MP2  &&  HF   &&&  M06 & B3LYP & BP   \\ \hline
def2-QZVPP   &   ref   & 0.29 &&  0.95 &&& 0.27 & 0.24  & 0.58 \\
def-TZVPP    &  0.09   & 0.30 &&  0.95 &&& 0.26 & 0.22  & 0.61 \\
def-TZVP     &  0.13   & 0.39 &&  0.94 &&& 0.27 & 0.24  & 0.59 \\
def-SVP      &  0.18   & 0.40 &&  0.98 &&& 0.32 & 0.28  & 0.62 \\ 
def-SV(P)    &   -     &  -   &&  0.98 &&& 0.35 & 0.32  & 0.67 \\ \hline
\hline \noalign{\smallskip}
\end{tabular}
\end{table}

In Table~\ref{tab.therm-error-bs} the basis set convergence is shown. 
The convergence with increasing basis sets is very shallow in all cases. 
That means, that economical basis sets are already sufficient to
compute the $\Delta G_{therm}$.  

\begin{table}
\caption{The RMS error in kcal/mol of $\Delta G_{therm}$ for different correlated methods with the def-TZVPP basis set compared to CCSD(T)/def2-QZVPP values and the scaling of computational costs with the system size.
\label{tab.therm-error-scaling} }
\smallskip
\begin{tabular}{lcc}  \hline
           &    RMS   & scaling\\ \hline
MP2        &   0.30   & $O(N^5)$  \\
SCS-MP2    &   0.28   & $O(N^5)$  \\ 
CCSD       &   0.18   & $O(N^6)$ \\
CCSD(T)    &   0.09   & $O(N^7)$ \\ \hline
\hline \noalign{\smallskip}
\end{tabular}
\end{table}

The performance of correlated methods in Table~\ref{tab.therm-error-scaling}
is compared to the computational scaling of the methods. Here, it can be seen that the
gain of going from cheaper to more expensive methods is systematic, but rather small.
DFT methods are in the same ballpark as on MP2 level, see Table~\ref{tab.error-func}.
DFT should be here the method of choice, because the vibrational frequency
computations can be become the time--determining step for larger molecules.

\begin{table}
\caption{Comparison of the performance of different density functionals with def-TZVP
by the RMS error in kcal/mol for $\Delta G_{therm}$.
\label{tab.error-func} }
\smallskip
\begin{tabular}{lcc}  \hline
        &  RMS   & functional class\\ \hline
BP      &  0.59  & GGA    \\
PBE     &  0.59  & GGA    \\ \hline
B3LYP   &  0.24  & HGGA   \\
PBE0    &  0.33  & HGGA   \\ \hline
TPSS    &  0.37  & MGGA   \\
M06-L   &  0.22  & MGGA   \\ \hline
M06     &  0.27  & HMGGA  \\
PW6B95  &  0.23  & HMGGA  \\ \hline
B97-D   &  0.43  & GGA-D3(BJ) \\ \hline
\hline \noalign{\smallskip}
\end{tabular}
\end{table}

Beside the methodological error, which is shown in the 
Tables~\ref{tab.therm-error-bs}, \ref{tab.therm-error-scaling}, and~\ref{tab.error-func}, 
there are additional errors, because the partition sums and chemical potentials 
are calculated within the rigid--rotor, harmonic--oscillator, and ideal gas approximations.
Taking into account better models is possible though computationally expensive. 
Such methods are 
not standard quantum chemistry application and the actual improvement is hard to quantify. 
In any case, we believe that the errors of $\Delta G_{therm}$ generally are the smallest and 
most well--behaved ones of the three building blocks.

\subsection{$ \Delta G_{solv}$}

The performance of COSMO-RS for the prediction of  
solvation free energies recently has been published in a work of 
Klamt and Diedenhofen\cite{dcosmors}.

\begin{figure}
\begin{center}
\includegraphics[width=8.9cm,angle=0]{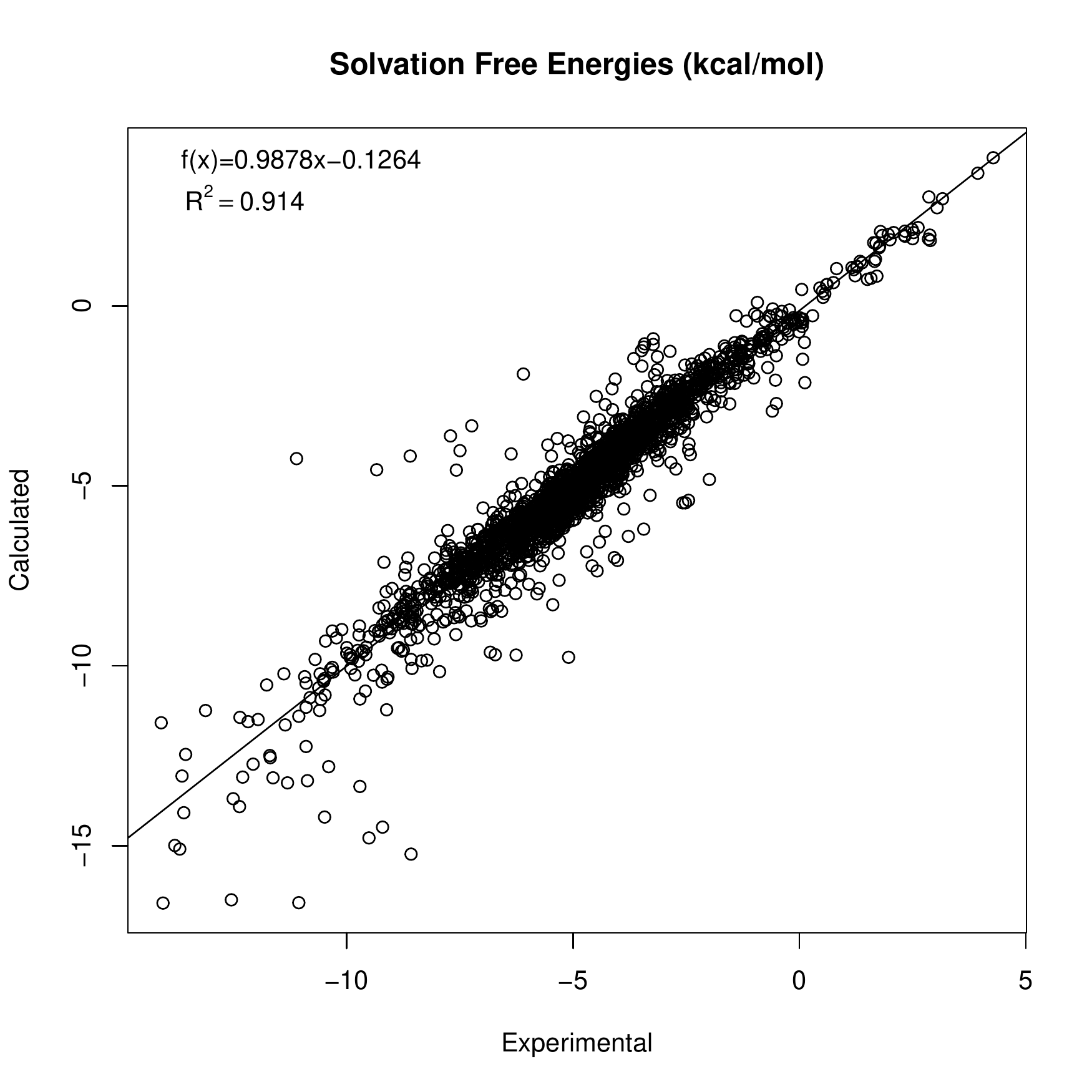} 
\end{center}
\caption{
Comparison experimental values of $\Delta G_{solv}$ with calculated ones by COSMO-RS theory.
The values are given in kcal/mol.
\label{fig.error-gsolv}}
\end{figure}

COSMO-RS free energies of solvation in Fig.~\ref{fig.error-gsolv}
have been calculated with the COSMO$therm$ program using the \texttt{BP\_TZVP\_C21\_0111}
parametrization and the standard conformer treatment outlined above.\cite{CT}
The experimental values were collected in the SM8 data set\cite{sm8},
which consists of 2346 neutral compounds at at 25\textdegree~C.
The mean unsigned error (MUE) for this test set is 0.42 kcal/mol
with an excellent correlation coefficient. The latest "high quality" 
\texttt{BP-TZVP} parameterization (\texttt{BP\_TZVP\_C30\_01701}) shows an MUE error 
of the same quantity. If, as discussed by Klamt and Diedenhofen,
data points with experimental or name ambiguities are removed, the MUE of the 
\texttt{BP-TZVP} level reduces to 0.39 kcal/mol. 
The prediction results for this cleaned up dataset 
improve to 0.36 kcal/mol MUE, if the most recent "best quality" \texttt{BP-TZVPD-FINE} level 
parametrization(\texttt{BP\_TZVPD\_FINE\_C30\_1701}) is used. 
A further improvement to 0.35 kcal/mol MUE is possible if the $\Delta G_{solv}$ predictions
are scaled by experimental vapor pressures, where they are available.

\section{Illustrative example}
\label{example}

As guinea pig we use the reaction acetaldehyde (ethanal) and hydrogen to
ethanol in toluene:
\begin{center}
CH$_3$CHO + H$_2$ $\rightarrow$ CH$_3$CH$_2$OH. 
\end{center}
Following the studies of Peters $et~al.$\cite{Peters08} 
we calculated the three building blocks of all products and reactants
on different level of theory.
The experimental data of the reaction free energy in the gas phase $\Delta G_{gas}$ 
and in solvent toluene $\Delta G_{soln}$ were taken from Ref. \onlinecite{Daubert88}
and Ref. \onlinecite{Adkins49}, respectively.

\begin{table}
\caption{Collection of $\Delta E_{gas}$, $\Delta G_{therm}$, and $\Delta G_{gas} = \Delta E_{gas} + \Delta G_{therm}$ 
for the gas phase acetaldehyde reduction on one specific level of theory.
Values are in kcal/mol.
\label{tab.ggas} }
\smallskip
\begin{tabular}{lccc}  \hline
               & $\Delta E_{gas}$ &  $\Delta G_{therm}$ & $\Delta G_{gas}$ \\ \hline 
BP/TZVP        & -20.0  &  15.8 & -4.2 \\
PBE/TZVP       & -21.5  &  15.8 & -5.7 \\ 
B3LYP/TZVP     & -20.3  &  16.0 & -4.2 \\
PBE0/TZVP      & -25.7  &  16.2 & -9.5 \\ 
TPSS/TZVP      & -18.4  &  15.8 & -2.5 \\
M06-L/TZVP     & -21.0  &  16.4 & -4.6 \\ 
M06/TZVP       & -22.0  &  16.6 & -5.4 \\
PW6B95/TZVP    & -22.2  &  16.1 & -6.1 \\ 
B97-D/TZVP     & -17.8  &  16.0 & -1.8 \\ \hline
MP2/TZVPP      & -23.8  &  16.2 & -7.6 \\
SCS-MP2/TZVPP  & -22.3  &  16.2 & -6.1 \\ 
CCSD/TZVPP     & -24.4  &  16.2 & -8.2 \\
CCSD(T)/TZVPP  & -23.4  &  16.1 & -7.3 \\ \hline
expt.          &        &       & -9.3 \\ \hline
\hline \noalign{\smallskip}
\end{tabular}
\end{table}

In Table~\ref{tab.ggas} results for the gas phase reaction for a selection of methods
is shown. For this reaction PBE0/TZVP is closest to the experimental value. However,
the DFT methods span a wider range of error than the WFT methods. 
The values of $\Delta E_{gas}$ vary about 8 kcal/mol,
while $\Delta G_{therm}$ only varies about 1 kcal/mol.
This encourages the use of two--level composite methods, which consist of  
single--point calculations on a high--level level of theory and geometry optimizations
and vibrational analysis with a cheaper method.
A sample of such composite methods are given in Table~\ref{tab.ggas-composite}.
It can be seen, that the use of a larger basis set for the single--point 
calculation has a much bigger effect than exchanging the method of geometry
optimization and vibrational analysis. 

\begin{table}
\caption{Collection of $\Delta E_{gas}$, $\Delta G_{therm}$, and $\Delta G_{gas}$
for the gas phase acetaldehyde reduction using a two--level composite scheme.
Values are in kcal/mol.
\label{tab.ggas-composite} }
\smallskip
\begin{tabular}{lccc}  \hline
                         & $\Delta E_{gas}$ &  $\Delta G_{therm}$ & $\Delta G_{gas}$ \\ \hline
MP2/TZVPP//BP/TZVP       & -23.7 & 15.8 & -7.8 \\ 
MP2/TZVPP//B3LYPP/TZVP   & -23.8 & 16.0 & -7.8 \\
MP2/TZVPP//TPSS/TZVP     & -23.6 & 15.8 & -7.8 \\
MP2/TZVPP//M06/TZVP      & -23.9 & 16.6 & -7.4 \\
MP2/QZVPP//BP/TZVP       & -24.2 & 15.8 & -8.4 \\ 
MP2/QZVPP//MP2/TZVPP     & -24.3 & 16.2 & -8.2 \\ \hline
CCSD(T)/TZVPP//BP/TZVP   & -23.3 & 15.8 & -7.4 \\
CCSD(T)/QZVPP//BP/TZVP   & -23.9 & 15.8 & -8.0 \\
CCSD(T)/QZVPP//MP2/TZVPP & -24.0 & 16.2 & -7.8 \\ \hline
expt.                    &       &      & -9.3 \\ \hline
\hline \noalign{\smallskip}
\end{tabular}
\end{table}


Using COSMO$therm$ with the \texttt{BP\_TZVP\_C30\_1701} parametrization yields -4.6 kcal/mol, 
and -3.9 kcal/mol with the \texttt{BP\_TZVPD\_FINE\_C30\_1701} parametrization. 
The differences between the two parametrizations is quite small in this case,
but the FINE parametrization should become superior for larger molecules, especially
for such that can form hydrogen bonds with the solvent.

The results for the free energy reaction in solution are collected in 
Table~\ref{tab.grxn}. The experimental values can be reproduced within a few kcal/mol. 

\begin{table}
\caption{Collection of $\Delta G_{soln} = \Delta G_{gas} + \Delta G_{solv}$
values in kcal/mol on different level of theory
for the acetaldehyde reduction in toluene.
For $\Delta G_{soln}1$ $\Delta G_{solv}$ from the \texttt{BP\_TZVP\_C30\_1701} parametrization
was used and for $\Delta G_{soln}2$ from the \texttt{BP\_TZVPD\_FINE\_C30\_1701} parametrization.
\label{tab.grxn} }
\smallskip
\begin{tabular}{lcc}  \hline
               & $\Delta G_{soln}1$ &  $\Delta G_{soln}2$  \\ \hline
BP/TZVP        & -8.8  & -8.1  \\
PBE/TZVP       & -10.3 & -9.6  \\
B3LYP/TZVP     & -8.8  & -8.1  \\
PBE0/TZVP      & -14.1 & -13.4  \\
TPSS/TZVP      & -7.1  & -6.4  \\
M06-L/TZVP     & -9.2  & -8.5  \\
M06/TZVP       & -10.0 & -9.3  \\
PW6B95/TZVP    & -10.7 & -10.0  \\
B97-D/TZVP     & -6.4  & -5.7 \\ \hline
MP2/TZVPP      & -12.2 & -11.5  \\
SCS-MP2/TZVPP  & -10.7 & -10.0  \\
CCSD/TZVPP     & -12.8 & -12.1  \\
CCSD(T)/TZVPP  & -11.9 & -11.2 \\ \hline
MP2/TZVPP//BP/TZVP       & -12.4 & -11.7 \\
MP2/TZVPP//B3LYPP/TZVP   & -12.4 & -11.7 \\
MP2/TZVPP//TPSS/TZVP     & -12.4 & -11.7 \\
MP2/TZVPP//M06/TZVP      & -12.0 & -11.3 \\
MP2/QZVPP//BP/TZVP       & -13.0 & -12.3 \\
MP2/QZVPP//MP2/TZVPP     & -12.8 & -12.1 \\ \hline
CCSD(T)/TZVPP//BP/TZVP   & -12.0 & -11.3 \\
CCSD(T)/QZVPP//BP/TZVP   & -12.6 & -11.9 \\
CCSD(T)/QZVPP//MP2/TZVPP & -12.4 & -11.7 \\ \hline
expt.          & \multicolumn{2}{c}{-10.4}  \\ \hline
\hline \noalign{\smallskip}
\end{tabular}
\end{table}

\section{Conclusion}
\label{conclusion}

The calculation of the Gibbs free energy of reactions in solution 
can be split into three parts according to Hess's law.
We have benchmarked the accuracy that can be achieved by different
level of theory separately for the three parts. 
The largest source for improvment for the total $\Delta G_{soln}$
is the choice of method for the calculation of $\Delta E_{gas}$. 

With the presented data recommendations for different levels can
be made and suitable computational workflow can be set up.
An adequate workflow can look like the following.

\begin{itemize}
\item TURBOMOLE computations
\begin{itemize}
\item Step 1a: \\
Optimize the structures of products and reactants as well 
as of the solvent molecule using BP/TZVP in the COSMO phase.
The resulting .cosmo files will be needed in step 5,
if the \texttt{BP-TZVP} level is used. 

\item Step 1b: \\
Calculate the BP/TZVPD single--point energy on the structures 
of step 1a in the COSMO phase
with a smooth radii--based isosurface cavity.
The resulting .cosmo files will be needed in step 5,
if the \texttt{BP-TZVPD-FINE} level is used.

\item Step 2a: \\
Optimize the structures of products and reactants 
in the gas phase.
The resulting .energy files will be needed in step 5,
if the \texttt{BP-TZVP} level is used.

\item Step 2b: \\
Calculate the BP/TZVPD single--point energy on the structures 
of step 2a in the gas phase
The resulting .energy files will be needed in step 5,
if the \texttt{BP-TZVPD-FINE} level is used.

\item Step 3: \\
Calculate the vibrational frequencies for the structures 
of step 2a using using BP/TZVP in the gas phase and afterwards 
the thermodynamic properties at 1 bar and the desired temperature (e.g. room temperature). 
The $\Delta G_{therm}$ contributions, also denoted chemical 
potential, to the thermodynamic cycle will be obtained here. 

\item Step 4: \\ 
Run single--point gas phase energy calculation for refinement of $\Delta E_{gas}$
for the structures of step 2a 
using MP2, SCS-MP2, or CCSD(T) with in a TZVPP or QZVPP basis, 
preferably the highest level which is feasible.

\end{itemize}

\item COSMO$therm$ computations

\begin{itemize}

\item Step 5: \\ 
Use the .cosmo and .energy files from step 1 and step 2 as input for a $\Delta G_{solv}$
calculation on \texttt{BP-TZVPD-FINE} level (recommended), or on \texttt{BP-TZVP} or 
\texttt{DMOL3-PBE} level if \texttt{BP-TZVPD-FINE} is not available. The $\Delta G_{solv}$ 
calculations of the reactant and product compounds should be done at infinite dilution 
in the solvent at the desired temperature using 1 bar gas to 1 mol solvent as reference state.

\end{itemize}
\end{itemize}

Adding up all the contributions according to Eq.~(\ref{e0}) will then yield the 
Gibbs free energy of reactions in solution $\Delta G_{soln}$.

\section*{Acknowledgments}

We thank Michael Diedenhofen for discussions and critically 
reading the manuscript.

\bibliography{refs} 

\end{document}